\documentstyle[12pt]{article}
\begin{document}

\pagestyle{empty}
\begin{flushright}
MADPH--95--921 \\
UCD--95--41 \\
December~1995 \\
\end{flushright}

\vspace{2cm}
\begin{center}

{\Large \bf Multiple Interactions in Two--Photon Collisions} \\
\vspace{1cm}
Manuel Drees$^a$ and Tao  Han$^b$ \\

\vspace{0.5cm}
{\it $^a$ Physics Department., University of Wisconsin, Madison, WI 53706,
USA} \\
{\it $^b$ Department of Physics, University of California, Davis, CA 95616,
USA} \\

\end{center}
\begin{abstract}
We compute cross sections for events where two pairs of partons scatter off
each other in the same $\gamma\gamma$ reaction, giving rise to at least 3
high--{\mbox{$p_T^{}$}} jets. Unlike in {\mbox{$p \bar p$}}\ collisions we
find the signal to lie well above the background from higher order QCD
processes. If the usual ``eikonaliztion" assumption is correct, the signal
should be readily observable at LEP2, and might already be detectable in data
taken at TRISTAN.
\end{abstract}

\vskip 0.5cm

\pagestyle{plain}
\clearpage
\setcounter{page}{1}
The successful predictions of cross sections for the production of hadronic
jets with high transverse momentum
{\mbox{$p_T^{}$}} is one of the most convincing tests
of perturbative QCD. The inclusive jet pair cross section can be
written as
\begin{equation} \label{e1}
\sigma (A B \rightarrow {\rm jets}) = \sum_{i,j,k,l} \int d x_1 f_{i|A}(x_1)
\int d x_2 f_{j|B}(x_2) \int_{p_{T,{\rm min}}} d p_T \frac {d \hat{\sigma} (ij
\rightarrow kl)} {d p_T}.
\end{equation}
Here $A$ and $B$ denote the projectiles (e.g. $p, \bar p$ or $\gamma$),
$i,j,k,l$ stand for parton species [(anti)quarks or gluons], $f_{i|A}(x)$ is
the distribution function of parton $i$ in hadron $A$, and the
{\mbox{$\hat{\sigma}$}} are the cross sections for the hard partonic
subprocesses. Eq.(\ref{e1}) describes the production of jets with $p_T^{} \geq
p_{T,{\rm min}}^{}$. Perturbative QCD is only applicable if the problem at
hand involves a sufficiently large momentum transfer, which implies $p_{T,{\rm
min}}^{} ({\rm parton}) \geq 1$ to 2 GeV. It has been suggested \cite{3} that
the production of these ``minijets" with ${\mbox{$p_T^{}$}} \sim 2$ GeV might
drive the observed increase of the total $pp$ and {\mbox{$p \bar p$}}\ cross
sections with energy.

Experiments at {\mbox{$p \bar p$}} colliders have only been able to detect
\cite{2} jets with ${\mbox{$p_T^{}$}}($jet$) \geq 5$ GeV. On the other hand,
experiments at TRISTAN \cite{4} and LEP \cite{5} have recently reconstructed
jets, produced in $\gamma\gamma$ collisions, with transverse momenta as low as
2 GeV. This allows to directly probe partonic processes down to the
theoretical cut--off on {\mbox{$p_T^{}$}}. Three distinct classes of reactions
contribute to the inclusive production of jets in $\gamma\gamma$ collisions
\cite{7}. This is due to the dual nature of the photon, which can interact
either ``directly", via its pointlike $\gamma q \bar q$ coupling, or as a
``resolved photon" \cite{8}, via its quark and gluon content. Here we are
interested in the ``twice resolved" contribution, where {\em both} incident
photons participate via their partonic content. Note that each resolved photon
gives rise to a remnant jet. Although the axes of these jets nearly coincide
with the beam holes, the TOPAZ collaboration has demonstrated \cite{9} that
their outer fringes can be detected. One can therefore experimentally isolate
a sample of twice resolved events.

The predicted jet cross sections obviously depend on the parton densities in
the photon $f_{i|\gamma}$, see eq.(\ref{e1}), which are at present only poorly
determined experimentally \cite{8}. Here we use the ``DG" parametrization of
ref.\cite{10}, which is in reasonable agreement with all data.
If {\mbox{$p_{T,{\rm min}}^{}$}} is
much smaller than the $\gamma\gamma$
center--of--mass energy {\mbox{$W_{\gamma \gamma}$}},
this parametrization predicts approximately
\begin{equation} \label{e2}
{\mbox{$\sigma (\gamma \gamma \rightarrow {\rm jets})$}}
\simeq 250 \ {\rm nb} \left( \frac {W_{\gamma \gamma}}{50 \ {\rm GeV}}
\right)^{1.4} \left( \frac {1.6 \ {\rm GeV}} {p_{T,{\rm min}}^{}}
\right)^{3.6}.
\end{equation}
Clearly the inclusive jet cross section grows very quickly with energy. In
contrast, total hadronic cross sections at high energy grow \cite{11} much
more slowly, $\propto s^{0.08}$, where
{\mbox{$\sqrt{s}$}} is the total center--of--mass
energy. This is true not only for hadron--hadron scattering, but also
describes the behaviour of the total $\gamma p$ cross section at HERA quite
well. Although the high--energy behaviour of the total $\gamma\gamma$
cross section
has not yet been measured, it seems reasonable to assume that it follows the
same asymptotic law, at least approximately.

If this is indeed the case, the jet cross section (\ref{e2}) exceeds the total
$\gamma\gamma$ cross section for ${\mbox{$W_{\gamma \gamma}$}} > 50$ to 100
GeV, assuming $p_{T,{\rm min}}^{} \simeq 1.6$ to 2.5 GeV as indicated by
analyses of multi--hadron production in  $\gamma\gamma$ scattering
\cite{4,5,8}. This apparent paradox is resolved once we realize that
eq.(\ref{e2}) describes an {\em inclusive} cross section, the definition of
which includes the average jet pair multiplicity:
\begin{equation} \label{e3}
{\mbox{$\sigma (\gamma \gamma \rightarrow {\rm jets})$}}
= \langle n_{\rm jet \ pairs} \rangle \sigma_{\rm tot} ( \gamma\gamma
\rightarrow {\rm hadrons}).
\end{equation}
Perturbative QCD thus predicts that the average number of jet pairs per
collision grows quickly with energy. This is true for both {\mbox{$p \bar p$}}
and $\gamma\gamma$ collisions. These additional jet pairs are thought to be
produced by scattering several pairs of partons off each other. In order to
describe such multiple interactions, one usually assumes them to occur
independently of each other, at least at fixed impact parameter. This leads to
the ``eikonalization" of (mini)jet cross sections \cite{3}, which allows a
successful description of total $pp$ and {\mbox{$p \bar p$}} cross sections.

A very similar model of multiple partonic scattering is part of the successful
PYTHIA generator for {\mbox{$p \bar p$}} events. This greatly improves the
description of the ``underlying event'', including some quite subtle features
\cite{15}. Similarly, a recent study by the H1 collaboration \cite{16} showed
that details of energy flow patterns in photoproduction events with large
transverse energy are quite well described by generators that include a model
of multiple interactions, while generators without this feature fail. These
analyses strongly indicate that multiple interactions do indeed occur in
nature, at a rate compatible with the eikonalization ansatz. However, studies
of this kind are very sensitive to details of the event generator, including
hadronization effects etc.; and the behaviour of the total cross section can
also be reproduced by quite different models. In order to decisively test
eikonalization one has to look for events that show unambiguous evidence for
multiple scattering, that is, events with two (or more) {\em reconstructed}
jet pairs.

Experiments at $pp$ and {\mbox{$p \bar p$}} colliders have searched for such
events. The AFS collaboration \cite{17} at the CERN ISR claimed to have found
a much larger signal than theoretically expected; however, at that time no
full calculation of the QCD backgrounds from $2 \rightarrow 4$ reactions was
available. The UA2 collaboration \cite{18} only quoted an upper limit on the
rate for events with multiple partonic interactions. Most recently, the CDF
collaboration \cite{19} found indications (at about $2 \sigma$ level) that
some 5\% of all 4--jet events with ${\mbox{$p_T^{}$}}($jet$) \geq 30$ GeV are
due to multiple interactions. In contrast, we find that $\gamma\gamma$
experiments should be able to test the eikonalization ansatz (modified for
reactions involving resolved photons \cite{12}) decisively, at the latest at
LEP2, and possibly even using data already taken at TRISTAN. Such a test would
have ramifications in many areas of collider physics. For example, the
``underlying event" in $pp$ and {\mbox{$p \bar p$}} collisions determines to
what extent hard leptons and photons remain isolated from hadronic debris,
which is a crucial issue when assessing backgrounds to ``new physics''
signals.

Since we assume partonic scatters to occur independently of each other, the
cross section for events with at least two pairs of high--{\mbox{$p_T^{}$}}
jets is proportional to the square of the single jet pair inclusive cross
section:
\begin{equation} \label{e5}
\sigma (\gamma\gamma \rightarrow \geq 2 \ {\rm jet \ pairs}) =
\left[ {\mbox{$\sigma (\gamma \gamma \rightarrow {\rm jets})$}}
\right]^2 / \sigma_0.
\end{equation}
For the DG parametrization and minimal partonic ${\mbox{$p_T^{}$}} = 1.6$ GeV,
within the eikonalization ansatz $\sigma_0 \simeq 300$ nb approximately
reproduces the ``universal" $s^{0.08}$ behaviour of the total cross section
\cite{14}; we therefore adopt this value as our standard choice. Since we
assume that the two partonic scatters do not ``know" of each other, energy and
momentum conservation have to hold for each scatter separately. Each partonic
interaction then gives rise to a pair of jets with equal and opposite
transverse momenta. Further, the distribution in the transverse opening angle
between the two pairs should be flat. These properties can be used to
distinguish between the signal from double parton scattering and the
background from higher order QCD processes.

In order to take the finite angular coverage of real detectors into account,
we only accept jets with rapidity $|y({\rm jet})| \leq y_{\rm cut}$.
We consider both the 3--jet signal $\sigma_{34}$
and the 4--jet signal $\sigma_{44}$, where the second subscript denotes the
total number of produced high--{\mbox{$p_T^{}$}} jets,
and the first subscript is the
number of accepted jets. In case of $\sigma_{34}$, we also require the missing
{\mbox{$p_T^{}$}},
defined as the negative vectorial sum of the transverse momenta of the
three accepted jets, to exceed {\mbox{$p_{T,{\rm min}}^{}$}}.
In the absence of jet energy smearing,
in signal events this missing {\mbox{$p_T^{}$}}
vector is equal and
opposite to the {\mbox{$p_T^{}$}}
vector of one of the three accepted jets, with the other
two jets being back--to--back to each other.
We require that all partons have separation $\Delta R = \sqrt{ \left( \Delta
\phi \right)^2 + \left( \Delta y \right)^2} \geq 0.75$ from each other.
To simulate detector resolution effects, we smear the energies of all
final state partons inside the acceptance region, assuming Gaussian
fluctuations with a relative error of 35\%; this number is characteristic for
the TOPAZ detector \cite{21}.

In Table 1 we show cross sections for signals and backgrounds at TRISTAN
(${\mbox{$\sqrt{s}$}}= 58$ GeV) and LEP2
(${\mbox{$\sqrt{s}$}} = 175$ GeV).
We have generated events with partonic {\mbox{$p_T^{}$}}
down to 1.6 GeV, but required ${\mbox{$p_T^{}$}}({\rm jet}) \geq 2.0$
(2.5) GeV for
TRISTAN (LEP2); in case of the 3--jet signal, the same cut has been applied on
the missing {\mbox{$p_T^{}$}}.
Our choices for $y_{\rm cut}$, 1.0 for TRISTAN and 1.75 for
LEP, have been inspired by the AMY and ALEPH detectors, respectively. The
signal calculation is based on eq.(\ref{e5}), while the background to
$\sigma_{44}$ has been calculated by adapting a computer code for 4--jet
production at {\mbox{$p \bar p$}}
colliders \cite{20}. The estimate of the background
contribution to $\sigma_{34}$ is a bit more complicated. The missing
transverse momentum is sensitive to fluctuations in the measured jet energies.
It is possible that events with only three hard partons produce a sufficient
amount of ``fake" missing {\mbox{$p_T^{}$}}
entirely due to such fluctuations. One
therefore has to deal with soft singularities when estimating the background
to the 3--jet signal from final states containing 4 partons. We used the
``poor man's shower" approach of ref.\cite{22} to regularize these
divergencies. Here one multiplies the squared matrix element for the $2
\rightarrow 4$ process with a function $f(p_{T,low}) = 1 - \exp(-p^2_{T,low} /
p^2_0)$, where $p_{T,low}$ is the smallest of the four transverse momenta. The
constant $p_0$ is chosen such that this calculation reproduces the 3--jet
cross section, {\em without} missing {\mbox{$p_T^{}$}}
requirement, as estimated from the
$2 \rightarrow 3$ matrix element. The idea is that the fourth, softest parton
then approximates the effects of initial and final state radiation. Finally,
all  $\gamma\gamma$
cross sections have been convoluted with effective photon spectra,
as determined by an experimental veto (``anti--tag") of events with outgoing
$e^\pm$ at angle $\theta > 5^{\circ}$ with respect to the beam pipes; this
ensures that the virtuality of both incident photons is small.

The first column of Table 1 shows signal and background after only the basic
acceptance cuts have been imposed. It is very encouraging to note that the
background is already smaller than the signal. In column 2 we in addition
require the jets to be pairwise back--to--back; in case of $\sigma_{34}$ this
also means that the missing {\mbox{$p_T^{}$}} vector must be back--to--back
with the unpaired jet. In order to allow for a measurement error in the
determination of the jet axes, as well as hadronization effects, we accept
pairs if $\cos \phi_{jj} \leq -0.9$ ($\phi_{jj} > 154^{\circ}$). This leaves
the signals unchanged (recall that we only fluctuate the jet energies, not
their angles), but reduces the backgrounds by about a factor of 2.5. Next we
impose a cut on the {\mbox{$p_T^{}$}}--balance of the paired jets, and of the
unpaired jet with the missing {\mbox{$p_T^{}$}} vector. To that end, we demand
\begin{equation} \label{e6}
\left| \vec{{\mbox{$p_T^{}$}}}(1) +
\vec{{\mbox{$p_T^{}$}}}(2) \right| \leq c \cdot \delta \cdot \max
\left( \left| \vec{{\mbox{$p_T^{}$}}}(1)\right|,
\left| \vec{{\mbox{$p_T^{}$}}}(2) \right| \right),
\end{equation}
where $\delta = 0.35$ is the jet energy resolution, and $c$ is a constant. The
choice $c = 1.5$ (column 3) does not reduce the signal significantly even
after smearing, whereas $c=1.0$ (column 4) reduces $\sigma_{34}$ by about 10\%
and $\sigma_{44}$ by about 20\%. However, the tighter cut reduces the
background even more, so we adopt it as standard from now on. After this cut,
we have reached a signal to noise ratio of at least 5 to 1 in all cases.

In the signal the distribution in
the transverse angle between the jet pairs should be flat, whereas the
background prefers some jets to be close to each other. In the last column of
table 1 we therefore in addition require this angle to be larger than
26$^{\circ}$ ($\cos \phi_{\rm pair} \leq 0.9$). This enhances the significance
of the 3--jet signal somewhat. However, since we have not attempted to model
the uncertainties in the determination of jet axes, we do not apply this last
cut in the numerical results presented below.

Table 1 shows that, for the given cuts, the 4--jet signal at TRISTAN is quite
marginal. Nominally the integrated luminosity of about 300 pb$^{-1}$ collected
by each of the three TRISTAN experiments would give rise to some 15 signal
events, on a background of a few events. However, this estimate does not yet
include a jet reconstruction efficiency, which could be significantly less
than unity for such soft jets. On the other hand, our calculation indicates
that the 3--jet plus missing {\mbox{$p_T^{}$}}
signal might be viable already at TRISTAN.
Even if the reconstruction efficiency is only 50\% per jet, we would still
expect about 15 signal events, on a background of 2 or 3 events.

In Fig.~1a,b we show the dependence of $\sigma_{34}$ (dashed) and
$\sigma_{44}$ (solid) at TRISTAN on {\mbox{$p_{T,{\rm min}}^{}$}}
and $y_{\rm cut}$. As expected
from eqs.(\ref{e2}) and (\ref{e5}), the signal depends very strongly on
{\mbox{$p_{T,{\rm min}}^{}$}}.
Fig.~1b shows that the 4--jet signal also depends
very sensitively on $y_{\rm cut}$; the signal would more than double if
$y_{\rm cut}$ could be increased from 1.0 to 1.25. The 4--jet background shows
basically the same dependence on $y_{\rm cut}$ as the signal. In case of the
3--jet rate, the background grows slightly more rapidly when $y_{\rm cut}$ is
increased; nevertheless the signal remains well above the background
everywhere.

In Fig.~2a,b we show the same distributions at LEP2 energy. Assuming an
integrated luminosity of 500 pb$^{-1}$, for the cuts of Table 1 we expect
several hundred signal events at least. Fig.~2a shows that the signal might
remain viable out to $p_{T,{\rm min}}^{} \simeq 4$ GeV. The background falls
less quickly with increasing {\mbox{$p_{T,{\rm min}}^{}$}} than the signal;
this trend is already visible in Fig.~1a. Nevertheless the signal to noise
ratio remains well above unity; the signal is essentially rate limited.
Fig.~2b shows that much of the advantage of LEP2 over TRISTAN is due to the
better angular coverage of the tracking system (which determines $y_{\rm
cut}$) of LEP detectors. On the other hand, the rapidity plateau for
${\mbox{$\sqrt{s}$}}=175$ GeV, $p_{T,{\rm min}}^{} = 2.5$ GeV is considerably
wider than for ${\mbox{$\sqrt{s}$}}=58$ GeV, $p_{T,{\rm min}}^{} = 2.0$ GeV;
for fixed $y_{\rm cut}$, the probability to have at least one jet outside the
acceptance region, and hence the ratio $\sigma_{34}/\sigma_{44}$, is therefore
much larger in Fig.~2b than in Fig.~1b.

Detection of the signals discussed here may require to change the criteria
used to experimentally define samples of two--photon events. Traditionally,
cuts on the visible invariant mass or visible energy of the entire event have
been used to suppress backgrounds from {\mbox{$e^+e^-$}}
annihilation. This may be
dangerous, since the signal cross section gets significant contributions from
events with ${\mbox{$W_{\gamma \gamma}$}} \geq
{\mbox{$\sqrt{s}$}}(e^+e^-)/2$ \cite{14}.
It seems much safer to cut on
the invariant mass of {\em only} the multi--jet system that forms the signal.
We find that almost no signal will be lost if one rejects events with $W \ge
{\mbox{$\sqrt{s}$}}(e^+e^-)/3$; in contrast, annihilation backgrounds
peak at $W \simeq {\mbox{$\sqrt{s}$}}(e^+e^-)$.

Let us briefly discuss the uncertainties of our predictions. Both signal and
backgrounds are of fourth order in $\alpha_s$. The results presented here are
therefore quite sensitive to the choice of momentum scale, and of the QCD
scale parameter $\Lambda$. We used the (average) jet transverse momentum
(before smearing) for the former, and a leading order value $\Lambda = 0.4$
GeV for the latter, as appropriate for the DG parametrization, assuming
$N_f=4$ active flavors. Our choice of structure function is a conservative
one; other popular parametrizations give similar or larger signal rates. The
main theoretical uncertainty comes from the eikonalization ansatz. As
discussed in ref.\cite{14}, it cannot be completely correct; eq.(\ref{e5}) as
written (slightly) violates energy conservation. We have corrected for this
using the rescaling method of ref.\cite{15}, which seems to work reasonably
well in {\mbox{$p \bar p$}}
collisions, but may not be applicable to $\gamma\gamma$ collisions.
Further, even if (mini)jet production in $\gamma\gamma$
collisions can be eikonalized,
it is not clear whether a single eikonal, as assumed in eq.(\ref{e5}), will be
sufficient; more complicated schemes have been suggested \cite{24}. We will
study these issues in a future publication.

We believe that our estimate of signal rates should be accurate
to within a factor of 2 to 3, {\em if} the basic ansatz is (approximately)
correct. In this case we find that detection of a clear signal for multiple
partonic scattering may be difficult at TRISTAN in the 4--jet channel; the
3--jet plus missing {\mbox{$p_T^{}$}}
signal looks more promising. Both signals should be
prominent at LEP2. Given the ambiguous outcome of searches for multiple
interactions at hadron colliders, this optimistic conclusion might come as a
surprise. However, $\gamma\gamma$
experiments have two crucial advantages. They can
reconstruct jets that are too soft to be detectable on top of the ``underlying
event" in {\mbox{$p \bar p$}}
collisions. Further, the ratio $\sigma({\rm jet})/\sigma_0$,
see eq.({\ref{e5}), is about ten times larger for $\gamma\gamma$
collisions than for {\mbox{$p \bar p$}}
collisions. $\gamma\gamma$ experiments might therefore allow a first {\em
direct} test of the eikonalization ansatz, which is a crucial ingredient of
current models of the ``underlying event" in hadronic collisions, and of
minimum bias physics in general.

\subsection*{Acknowledgements}
We thank Walter Giele for sending us a computer code based on the results of
ref.\cite{20}. The work of M.D. was supported in part by the U.S. Department
of Energy under grant No. DE-FG02-95ER40896, by the Wisconsin Research
Committee with funds granted by the Wisconsin Alumni Research Foundation, as
well as by a grant from the Deutsche Forschungsgemeinschaft under the
Heisenberg program. T.H. was supported in part
by DOE under grant DE--FG03--91ER40674.

\noindent

\clearpage
\noindent
{\bf Table 1:} Signal and background cross sections (in pb), as well as their
ratios, at TRISTAN and at LEP2. In the first column only the basic acceptance
cuts on the transverse momentum, rapidity of the jets and on the jet
separation ($\Delta R > 0.75$) have been applied. In the second column we in
addition require jets to be pairwise back--to--back, as described in the text,
and in columns 3 and 4 we also impose the cut (\ref{e6}) with $c=1.5$ and 1.0,
respectively. Finally, in column 5 we further impose a cut on the
transverse opening angle $\phi_{\rm pair}$ between jet pairs.

\begin{center}
\begin{tabular}{|c||c|c|c|c|c|}
\hline
in pb
& basic & back--to--back & $c=1.5$& $c=1.0$ & $\cos \phi_{\rm pair}\leq 0.9$
\\ \hline
\multicolumn{6}{|c|}{a) TRISTAN: $\sqrt{s}=58$ GeV,
$p_{T,{\rm min}}^{} =2.0$ GeV,
$y_{\rm cut}=1.0$} \\
\hline
$\sigma_{44}$(S) & 0.072 & 0.072 & 0.070 & 0.053  & 0.046 \\
$\sigma_{44}$(B) & 0.047 & 0.017 & 0.015 & 0.0065 & 0.0057 \\
S/B              & 1.7   & 4.2   & 4.7   & 8.2    & 8.1 \\
\hline
$\sigma_{34}$(S) & 0.52  & 0.52  & 0.51  & 0.46 & 0.42 \\
$\sigma_{34}$(B) & 0.28  & 0.13  & 0.12  & 0.082& 0.055 \\
S/B         & 1.9  & 4.0   & 4.3   & 5.6  & 7.6 \\
\hline
\multicolumn{6}{|c|}{b) LEP2: $\sqrt{s}=175$ GeV,
$p_{T,{\rm min}}^{}=2.5$ GeV,
$y_{\rm cut}=1.75$} \\
\hline
$\sigma_{44}$(S) & 3.6 & 3.6  & 3.4  & 2.5  & 2.2 \\
$\sigma_{44}$(B) & 1.8 & 0.59 & 0.49 & 0.20 & 0.17 \\
S/B              & 2.0& 6.1  & 6.9  & 12.5 & 12.9 \\
\hline
$\sigma_{34}$(S) & 15.4 & 15.4 & 15.2 & 13.4 & 10.8 \\
$\sigma_{34}$(B) & 5.5  & 2.2  & 2.1  & 1.3  & 0.90 \\
S/B              & 2.8 &  7.0 & 7.2  & 10.3 & 12.0 \\
\hline
\end{tabular}
\end{center}


\clearpage
\begin{thebibliography}{99}

\bibitem{3}
D. Cline, F. Halzen and J. Luthe, Phys. Rev. Lett. {\bf 31}, 491 (1973); L.
Durand and H. Pi, Phys. Rev. Lett. {\bf 58}, 303 (1987); M.M. Block, F. Halzen
and B. Margolis, Phys. Rev. {\bf D45}, 839 (1992).

\bibitem{2}
UA1 collab., C. Albajar et al., Nucl. Phys. {\bf B309}, 405 (1988).

\bibitem{4}
TOPAZ collab., H. Hayashii et al., Phys. Lett. {\bf B314}, 149 (1993);
AMY collab., B.J. Kim et al., Phys. Lett. {\bf B325}, 248 (1994).

\bibitem{5}
DELPHI collab., P. Abreu et al., Z. Phys. {\bf C62}, 357 (1994).

\bibitem{7}
S.J. Brodsky, T.A. DeGrand, J.F. Gunion and J. Weis, Phys. Rev. Lett. {\bf
41}, 672 (1978), and Phys. Rev. {\bf D19}, 1418 (1979); K. Kajantie and R.
Raitio, Nucl. Phys. {\bf B159}, 528 (1979).

\bibitem{8}
For a review, see M. Drees and R.M. Godbole, J. Phys. {\bf G21}, 1559 (1995).

\bibitem{9}
TOPAZ collab., M. Iwasaki et al., Phys. Lett. {\bf B341}, 99 (1994).

\bibitem{10}
M. Drees and K. Grassie, Z. Phys. {\bf C28}, 451 (1985).

\bibitem{11}
A. Donnachie and P.V. Landshoff, Phys. Lett. {\bf B296}, 227 (1992).

\bibitem{15}
T. Sj\"ostrand and M. van Zijl, Phys. Rev. {\bf D36}, 2019 (1987).

\bibitem{16}
H1 collab., S. Aid et al., DESY report 95--219, hep--ex 9511012.

\bibitem{17}
AFS collab., T. Akesson et al., Z. Phys. {\bf C34}, 163 (1987).

\bibitem{18}
UA2 collab., J. Alitti et al., Phys. Lett. {\bf B268}, 145 (1991).

\bibitem{19}
CDF collab., F. Abe et al., Phys. Rev. {\bf D47}, 4857 (1993).

\bibitem{12}
J.C. Collins and G.A. Ladinsky, Phys. Rev. {\bf D43}, 2847 (1991);
J.R. Forshaw and J.K. Storrow, Phys. Lett. {\bf B278}, 193 (1992).

\bibitem{14}
M. Drees, talk given at the {\it Third Workshop on TRISTAN Physics at High
Luminosities}, KEK, Tsukuba, November 1994, KEK Proceedings 95--6, eds. H.
Sagawa et al.

\bibitem{21}
H. Hayashii, private communication.

\bibitem{20}
F.A. Berends, W.T. Giele and H. Kuijf, Phys. Lett. {\bf B232}, 266 (1989);
F.A. Berends and H. Kuijf, Nucl. Phys. {\bf B353}, 59 (1991).

\bibitem{22}
V. Barger and R.J.N. Phillips, Phys. Rev. Lett. {\bf 55}, 2752 (1985).

\bibitem{24}
G.A. Schuler and T. Sj\"ostrand, Nucl. Phys. {\bf B407}, 539 (1993); K. Honjo,
L. Durand, R. Gandhi, H. Pi and I. Sarcevic, Phys. Rev. {\bf D48}, 1048
(1993).

\end{thebibliography}
\end{document}